\documentclass[manuscript,screen,authorversion, review=false]{acmart}

\renewcommand\footnotetextcopyrightpermission[1]{} 
\usepackage{longtable}

\AtBeginDocument{%
  \providecommand\BibTeX{{%
    \normalfont B\kern-0.5em{\scshape i\kern-0.25em b}\kern-0.8em\TeX}}}

\setcopyright{rightsretained}
\copyrightyear{2024}
\acmYear{2024}
\acmDOI{none}

\begin{document}

\title{Expanding Horizons in HCI Research Through LLM-Driven Qualitative Analysis}

\author{Maya Grace Torii}
\affiliation{%
  \institution{\\ R\&D Center for Digital Nature, University of Tsukuba}
  \streetaddress{1-2 Kasuga}
  \city{Tsukuba}
  \state{Ibaraki}
  \country{Japan}
  \postcode{305-0018}
}
\email{toriparu@digialnature.slis.tsukuba.ac.jp}
\orcid{0000-0003-4025-9212}

\author{Takahito Murakami}
\email{takahito@digialnature.slis.tsukuba.ac.jp}
\orcid{0000-0003-2077-9747}
\affiliation{%
  \institution{\\ R\&D Center for Digital Nature, University of Tsukuba}
  \streetaddress{1-2 Kasuga}
  \city{Tsukuba}
  \state{Ibaraki}
  \country{Japan}
  \postcode{305-0821}
}

\author{Yoichi Ochiai}
\affiliation{%
  \institution{R\&D Center for Digital Nature, University of Tsukuba}
  \streetaddress{1-2 Kasuga}
  \city{Tsukuba}
  \state{Ibaraki}
  \country{Japan}
  \postcode{305-0821}
}
\email{wizard@slis.tsukuba.ac.jp}
\orcid{0000-0002-4690-5724}

\renewcommand{\shortauthors}{Torii et al.}

\begin{abstract}
How would research be like if we still needed to "send" papers typed with a typewriter? Our life and research environment have continually evolved, often accompanied by controversial opinions about new methodologies. In this paper, we embrace this change by introducing a new approach to qualitative analysis in HCI using Large Language Models (LLMs). We detail a method that uses LLMs for qualitative data analysis and present a quantitative framework using SBART cosine similarity for performance evaluation. Our findings indicate that LLMs not only match the efficacy of traditional analysis methods but also offer unique insights. Through a novel dataset and benchmark, we explore LLMs' characteristics in HCI research, suggesting potential avenues for further exploration and application in the field.
\end{abstract}






\maketitle

\section{Introduction}
\textbf{As HCI researchers, our expertise lies in exploring human-computer interactions, not in laborious qualitative data coding.}
Qualitative research in Human-Computer Interaction (HCI) has traditionally been constrained by the extensive time, effort, and financial resources required for coding, categorizing, and interpreting large volumes of data. Traditional methods, despite their rigor, often face challenges in scaling and maintaining consistency across diverse contexts. The advent of Natural Language Processing (NLP) technologies has been offering some relief by automating aspects of these processes assisted in creating codebooks or categorizing data~\cite{Abram2020}.

A significant shift towards leveraging Large Language Models (LLMs) has been observed since the introduction of ChatGPT in 2022. The inherent strength of LLMs in processing and providing insights on substantial language data suggests a promising avenue for HCI research~\cite{Hämäläinen2023}. Studies have begun exploring the potential of LLMs in aiding qualitative analysis, particularly in tasks like codebook and categorizing~\cite{Tai2023, Bano2023}. However, these approaches often see NLP and LLMs as mere supplements to human effort, focusing on data compilation rather than interpretation, indicating an opportunity for further automation. As an exception, Byun et al.(2023) have made initial forays into using LLMs for generating discussions from raw qualitative data, although their approach lacks quantitative evaluation~\cite{Byun2023}.

In this paper, we propose a method for conducting qualitative analysis in HCI research using LLMs and evaluate the output by comparing it with the published paper. We aim to not only process and analyse datasets but also to enhance the reproducibility and scalability of qualitative research. This approach overcome the limitations of human-intensive methods, offering a pathway to handle vast amounts of information more efficiently. Specifically, we explored a quantitative evaluation of our LLMs system using a dataset comprising three published papers from the CHI and HCI domains, along with their raw questionare data. This leads to two significant contributions:

\begin{itemize}
\item Discover the characterization of LLM-driven qualitative analysis,
\item The first benchmark is proposed for evaluating LLMs performing qualitative analysis.
\end{itemize}

Our work suggests a potential direction for the future, where LLM-driven qualitative analysis in HCI research becomes a norm, ensuring high reproducibility, transparency, and scalability. This shift could revolutionize the way qualitative research is conducted, making it more accessible and efficient, and opening new horizons in understanding HCI.
\section{Related Works}
\subsection{Traditional Qualitative Methods in HCI}
Qualitative data analysis in human computer interaction(HCI) has traditionally relied on methods such as grounded theory, phenomenology, and ethnography~\cite{Rooksby2014,Dag2013,Anderson1994}. These methods have offered substantial insights into HCI, leveraging the strength of quantitative research allowing the researcher to familiarize themselves with the problem or concept under study, and potentially generate testable hypotheses~\cite{Olson2014,Golafshani2003}. However, they are also known for being labor-intensive, placing a significant demand on resources and time~\cite{Marathe2018,Gao2023}. With the advancements in computing and technology, efforts have been made to mitigate these labor-intensive tasks and to increase the transparency of qualitative evaluations. One approach is the use of codebooks in the qualitative analysis process. Codebooks provide a systematic and standardized method for categorizing and encoding the data~\cite{Adams2008}. This allows for a quantitative evaluation of qualitative data, which can serve to elucidate patterns and themes that might not be readily apparent through a purely qualitative analysis~\cite{Reyes2023}. Furthermore, the use of codebooks can increase the reliability of the data analysis process, as it helps to ensure consistency and accuracy in data coding~\cite{Weston2001}.

\subsection{Integration of NLP in Qualitative Analysis}
Advancements in NLP have significantly influenced qualitative analysis. Research comparing traditional qualitative text analysis, NLP, and their combination highlights the strengths and limitations of each approach. While NLP efficiently identifies key themes, it may lack the depth for nuanced context, a gap filled by qualitative methods~\cite{Guetterman2018}. The synergy of NLP and qualitative techniques offers more comprehensive analysis, enhancing the research process with efficient coding and validation of findings~\cite{Guetterman2018,Gao2023}. Further advancements focus on improving reliability, transparency, and deeper insights in qualitative analysis~\cite{Kaufmann2020,Leeson2019,Lennon2021}. LDA's application in diverse contexts, like analyzing social media during critical events, demonstrates NLP's potential in automating data interpretation~\cite{Zhou2021,Guven2021}. Additionally, evaluations of these NLP-based qualitative analysis methods by their users have been conducted, indicating their growing importance in research methodologies~\cite{Marathe2018}.

\subsection{Emergence of LLMs in Qualitative Research}
Since the emergence of ChatGPT in 2022, research has increasingly focused on leveraging LLMs for tasks in qualitative analysis such as codebook categorization. Reserches of Tai et al.(2023) and Bano et al.(2023) explore using LLMs to augment traditional coding methods in qualitative research~\cite{Tai2023,Bano2023}. Tai et al.(2023) proposes a methodology using LLMs(ChatGPT3.5) for deductive coding, comparing its efficacy against traditional human coding methods. Bano et al.(2023) examines the comprehension abilities of humans and LLMs(ChatGPT3.5 and GPT-4), analyzing their classification and reasoning capabilities in a study involving Alexa app reviews. These studies suggest the potential for effective human-LLM collaboration in qualitative research, emphasizing the need for continuous evaluation of LLMs' role in research practices.

Most current applications of NLP and LLMs in qualitative analysis, while easing the manual workload, still require substantial human involvement for interpreting data and generating insights. Studies of XiaoZiang et al.(2023) have shown that while AI tools can assist in coding, they cannot fully replace the nuanced understanding needed for data interpretation~\cite{Xiao2023}. Byun et al.(2023), however, have experimented with using LLMs to autonomously generate discussions from qualitative data without relying on a codebook, aiming for more independent and insightful analysis\cite{Byun2023}. This marks a significant step towards more autonomous analytical techniques in qualitative research.


\section{Method}
\subsection{Experiment Procedure}
This experimental setup includes the use of two different LLMs: GPT-4 of openAI and Llama 13B from meta\cite{OpenAI2023,Touvron2023}. To assess the capabilities of Large Language Models (LLMs) in qualitative data analysis, we have chosen the Sentence-BART(SBART)~\cite{Reimers2019} as our primary tool. These scores are derived from the comparison between each LLMs model's output and the original CHI papers. This quantitative analysis will allow us to evaluate the effectiveness of each LLM in understanding the nuances of qualitative research data and replicating. SBART is a pretrained BART which derives semantically meaningful sentence embeddings to be compare using cosine-similarity. This approach allows us to quantitatively compare the output of LLMs against original texts, providing a measurable metric of analysis quality. The primary objective of these experiments is to calculate average cosine-similarity. For each selected CHI paper text excerpt, we generate five iterations of output to see average cosine-similarity.

\subsection{Dataset}
Our study created a dataset comprising three CHI papers. These papers were selected based on the presence of qualitative evaluation related to open questions. We specifically focus on sections of these papers that describe the findings of open-question evaluations. The dataset we created includes:
\begin{itemize}
\item Summary: formatted summaries of the selected CHI papers.
\item Raw data: data pertaining to the open-ended questions and their responses.
\item Chosen passages from papers: set of passages that contains qualitative analysis findings and the preceding paragraph.
\end{itemize}
The summary were generated by the original published paper and GPT-4 assistant with summary prompt. Raw data was downloaded and organised in a program readable form with open ended question's questionnaire and answers from participants in a set. Passages from papers were extracted in the decision of research conductors if there are any information and findings from open-ended question contained. The containing main passage and the preceding paragraph was stored as data.
This dataset serves multiple purposes: it provides qualitative data to challenge and test the analysis capabilities of the LLMs, and it can be repurposed for other research or used as a benchmark for system improvements.

\subsection{Prompts}
\begin{figure}[thpb]
  \centering
  \includegraphics[width=\textwidth]{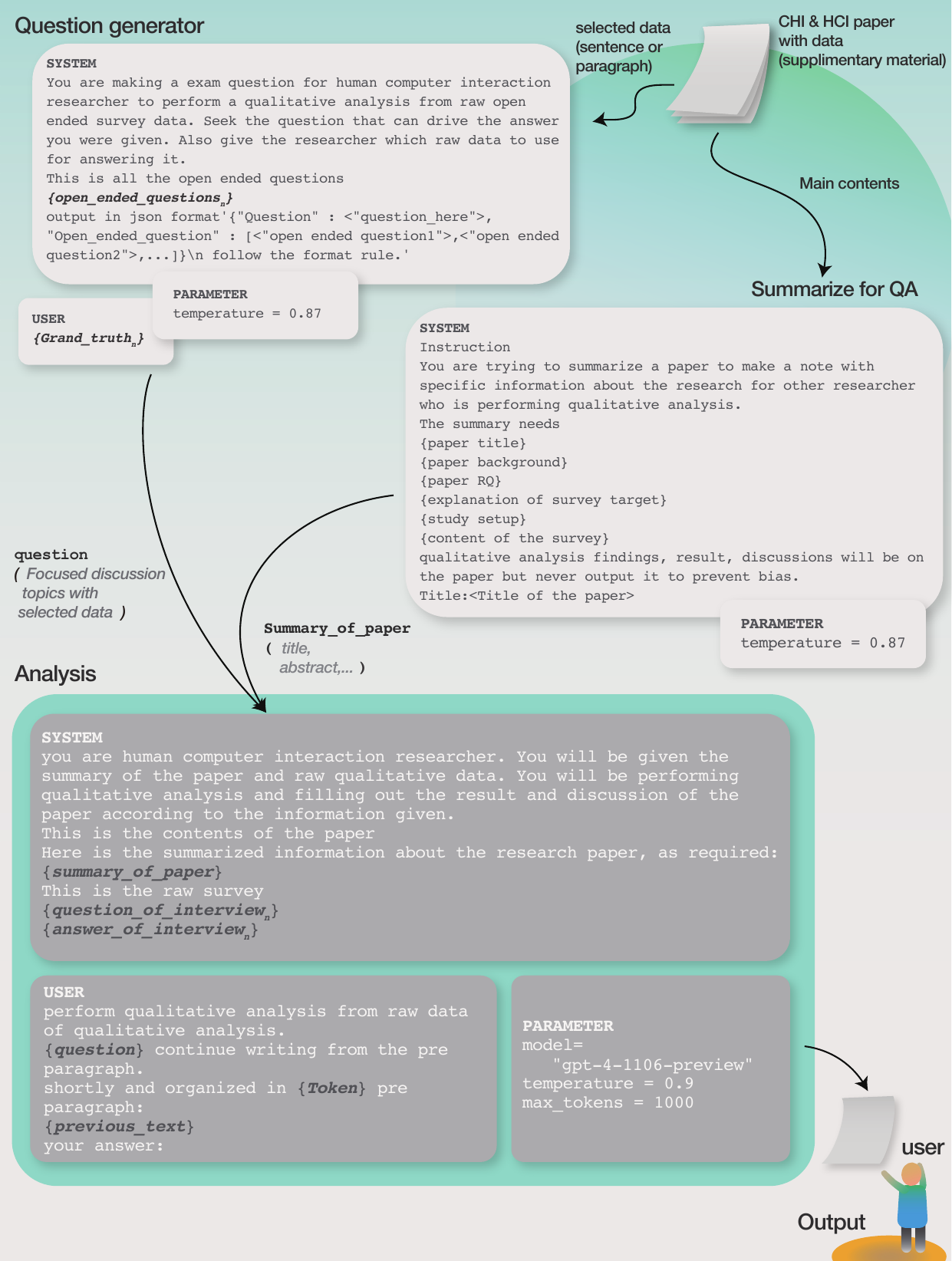}
  \caption{Three prompts used in the qualitative analysis LLMs system and the evaluation system.}
  \Description{prompts}
  \label{fig:prompt}
\end{figure}

To execute the experiment, three prompt system were prepared, analysis prompt, summarise for QA prompt, question generator prompt (Fig.~\ref{fig:prompt}).

\subsubsection{Analysis} The qualitative analysis prompt is a pivotal component of our system, designed to facilitate an autonomous approach to qualitative analysis by LLMs, distinct from traditional human-executed methods. Instead of dividing the task into multiple sub-tasks typically handled by human researchers, we integrated the process into a single, comprehensive prompt. Input require a summary of the research paper, rather than the full text, providing a concise context for the LLM. This approach ensures that the LLM can effectively handle extensive raw data by focusing on the essence of the research. Essential raw data necessary for analysis, the preceding paragraph of the text being generated, specific themes or research questions to be addressed in the output and The desired length of the output, defined in terms of tokens is required. 
This qualitative analysis is constructed for openAI model upper than GPT-4.

\subsubsection{Paper summary} The Paper Summary prompt is designed to create summaries of research papers, focusing on essential information while excluding findings and discussions to avoid bias. This prompt is specifically programmed to work with OpenAI's GPT-4 1106 preview model, chosen for its ability to process long texts effectively. By generating these summaries, we provide a streamlined and relevant input for our LLM based qualitative analysis, ensuring that the analysis is concentrated on the most pertinent aspects of each paper.

\subsubsection{Question generator} The Question Generation prompt is designed for verifying our LLM's qualitative analysis. It works by creating questions that require specific answers drawn from the original paper. This prompt takes the given answers and generates corresponding questions, helping us evaluate the LLM’s understanding of the data. It also identifies relevant raw data necessary for these questions.

\section{Result}
\subsection{SBART Score Result}
\begin{figure}[thbp]
  \centering
  \includegraphics[width=\textwidth]{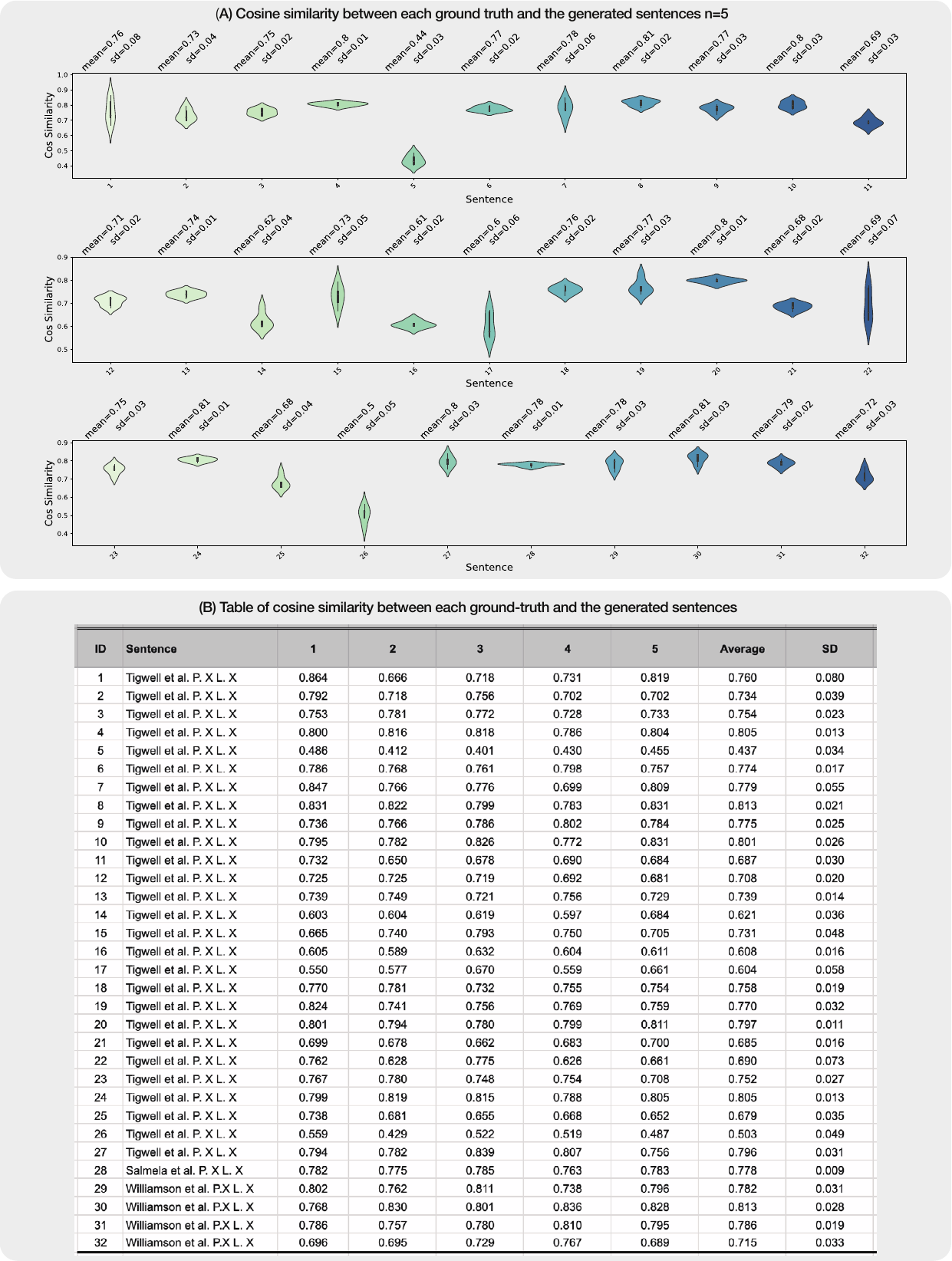}
  \caption{Cosine similarity results from SBERT. (A) violin plotted, (B) showing the relationship between the paper and the data. ID 1-27~\cite{10.1145/3313831.3376267}, 28~\cite{10.1145/3290605.3300732}, 29-32~\cite{10.1145/3290605.3300310}}
  \Description{Cosine similarity results from SBERT. (A) violin plotted, (B) showing the relationship between the paper and the data. ID 1-27~\cite{10.1145/3313831.3376267}, 28~\cite{10.1145/3290605.3300732}, 29-32~\cite{10.1145/3290605.3300310}}
  \label{resulttable}
\end{figure}

The analysis of qualitative analysis findings generated by the LLMs, specifically GPT-4 in this study, was quantitatively evaluated using SBART. The results indicate a noteworthy consistency in the outputs, as evidenced by the small standard deviations (SD < 0.02) in most cases. However, there were notable exceptions where lower cosine-similarity were observed (Fig.~\ref{resulttable}A, B).

The lowest scores were observed in specific outputs, notably in the case of Tigwell et al~\cite{10.1145/3313831.3376267}., where items 5 and 26 scored 0.437 and 0.503 respectively (Fig.~\ref{resulttable}B). Additionally, five other items demonstrated average scores below 0.7, indicating a certain pattern in lower-scoring outputs. These lower scores were predominantly found in sections where numerical data from coding results were presented, such as "in private (43 participants) and public (32) contexts...". Moreover, outputs incorporating direct participant responses as key frames, like P38’s statement about emoji selection challenges due to visual impairment, also tended to yield lower scores.
These findings suggest that while cosine-similarity provides a consistent measure of the LLM's output quality, certain types of content, particularly those involving numerical data or direct participant quotes, may pose challenges. 
\section{Discussion for this research}
\subsection{Limitation}
Our study encountered several limitations to be acknowledged and addressed in future research.

First there are limitations in the system of token size, which vary depending on the model. For instance, models like Llama2 have a token limit of approximately 4000, posing challenges in accommodating extensive data required for comprehensive qualitative analysis. This limitation becomes particularly critical for our qualitative analysis prompt, where reducing the size of raw data without compromising its integrity is not feasible. Consequently, the volume of raw data emerges as a bottleneck, potentially hindering the effectiveness of the analysis system. Adapting to these constraints requires careful consideration of data selection and processing strategies to ensure the system operates within the model's capacity while still providing meaningful analysis.

As highlighted by Julian Ashwin et al., the errors made by LLMs in coding and annotating qualitative data are not random but may be influenced by inherent biases in the models~\cite{Ashwin2023}. This necessitates a deeper investigation into the nature of these biases and their implications for qualitative analysis.

The GPT models of openAI remain largely opaque in their operation. With OpenAI frequently updating their specifications, the performance and outcomes of these models are subject to change. This underscores the need for continuous vigilance and adaptability in methods involving rapidly evolving LLMs.

\subsection{Toward HCI with LLMs}
As we investigate LLMs in scientific research, especially in qualitative analysis, several challenges and considerations emerge that warrant further discussion and exploration.

Our current method of comparing SBART scores primarily focuses on cosine similarity, which might not fully capture the essence of textual meanings and nuances. Cosine similarity compares word frequencies rather than the semantic integrity of the text, raising questions about its effectiveness in evaluating the quality of qualitative analysis.
However, the ultimate objective in qualitative research is not merely to replicate existing papers but to produce objective, persuasive, and significant results. Therefore, there's a need to explore better benchmarks to more accurately evaluate LLMs effectiveness in qualitative analysis.

There is a pressing need to define what constitutes sufficient qualitative analysis and to develop better benchmarks that reflect the demands of significant qualitative research. This includes devising improved methods for performing qualitative analysis and expanding the diversity and depth of datasets used. Future research should aim to establish more robust and nuanced evaluation criteria that go beyond mere textual similarity, capturing the true essence of qualitative insights.
\section{Conclusion}
This paper presented a method utilizing LLMs for qualitative analysis in HCI research, introducing a quantitative evaluation framework with SBART cosine similarity. The study demonstrated the potential of LLMs in processing qualitative data, provided a novel benchmark for evaluating their performance. The output generated by the LLMs in qualitative analysis suggest that while LLMs can yield results and discussions similar to traditional methods, they also offer unique approaches to qualitative analysis. These insights gained from understanding the characteristic of LLM-driven qualitative analysis, facilitated by the new dataset and benchmark, contribute to the ongoing discourse on the application of LLMs in research. These findings highlight areas for future investigation and development, of LLMs in qualitative research.

\begin{acks}
Great guardian of LLM server Naruya Kondo. The manuscript was drafted using OpenAI ChatGPT and GPT-4. The AI generated text was read, revised and proofed carefuly by the authors.
\end{acks}

\bibliographystyle{ACM-Reference-Format}
\bibliography{main-bibliography}


\end{document}